\documentclass{aa} 
\usepackage{epsfig,latexsym,graphicx,epsf}
\usepackage{natbib}
\usepackage{longtable} 
\usepackage{amsmath}

\def\lsim{\raise0.3ex\hbox{$<$}\kern-0.75em{\lower0.65ex\hbox{$\sim$}}}
\def\gsim{\raise0.3ex\hbox{$>$}\kern-0.75em{\lower0.65ex\hbox{$\sim$}}}

\begin{document}
\title{SNOC: a Monte-Carlo simulation package for high-z supernova observations}

\author{ A.~Goobar     \inst{1} \and
        E.~M\"ortsell  \inst{1} \and
        R.~Amanullah   \inst{1} \and
        M.~Goliath     \inst{2} \and
        L.~Bergstr\"om \inst{1} \and 
        T.~Dahl\'en      \inst{3} }
\authorrunning{Goobar et al.}

\offprints{A. Goobar, ariel@physto.se}

\institute{
Physics Department, Stockholm University, 
SCFAB, S-106 91 Stockholm, Sweden \and 
Swedish Defence Research Agency (FOI), S-172 90 Stockholm, Sweden \and
Stockholm Observatory, SCFAB, S-106 91 Stockholm, Sweden}

\authorrunning{Goobar et al.}

\date{Received ...; accepted ...}

\abstract{We present a Monte-Carlo package for
simulation of high-redshift supernova data, SNOC. Optical and
near-infrared photons from supernovae are ray-traced over cosmological
distances from the simulated host galaxy to the observer at Earth. The
distances to the sources are calculated from user provided
cosmological parameters in a Friedmann-Lema\^{\i}tre universe,
allowing for arbitrary forms of ``dark energy''.  The code takes into
account gravitational interactions (lensing) and extinction by dust,
both in the host galaxy and in the line-of-sight. The user can also
choose to include exotic effects like a hypothetical attenuation
due to photon-axion oscillations.  SNOC is primarily useful for
estimations of cosmological parameter uncertainties from studies of
apparent brightness of Type Ia supernovae vs redshift, with special
emphasis on potential systematic effects.  It can also be used to
compute standard cosmological quantities like luminosity distance,
lookback time and age of the universe in any Friedmann-Lema\^{\i}tre
model with or without quintessence.   } \maketitle

\section{Introduction}

The study of brightness of Type Ia supernovae (SNe) at high redshifts
has become one of the most important tools in observational cosmology
over the last few years
\citep{GoobarPerlmutter1995,Perlmutter1999,Riess1998,Riess2001},
giving the first direct observational evidence for a presently accelerating
universe.
With the increased capabilities of observing very high-redshift SNe,
more comprehensive analysis tools are needed \citep[see, e.g.,][]{sn97ff}.

In this work we describe a Monte-Carlo simulation package called {\em
The SuperNova Observation Calculator} (SNOC) which produces synthetic
samples of SN observations that can be used to estimate the accuracy
of the magnitude-redshift method for measuring cosmological parameters
as well as quantifying the possibility of constraining extragalactic dust
properties \citep{dust} or investigating the matter distribution in
the Universe \citep{compact}.  For example, SNOC can be used to
quantify the deviations in the Hubble diagram of Type Ia SNe
due to gravitational lensing or dust
extinction along the line-of-sight.  
Tools are also provided to compute the possible
contamination of an observational sample by
 core collapse SNe and the likelihood of observation
of multiple (lensed) images from individual high-$z$ SNe
\citep{Goobar:2002}.

\section{The program}

\begin{figure}
\includegraphics[width=\hsize]{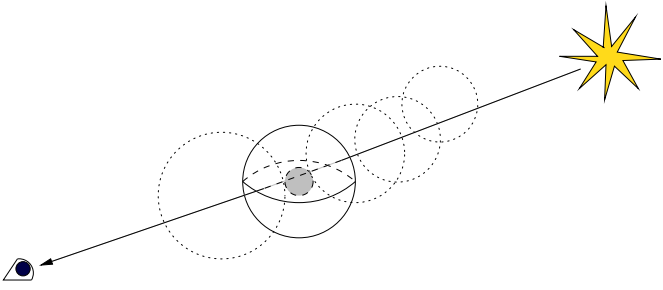}
\caption{Schematic view of the ray-tracing technique used in SNOC. The
light-beam is traced back-wards in time, from the observer to the host
galaxy taking into account the non-homogeneity of the universe and possible
intervening matter.}
\label{fig:schematic}
\end{figure}

SNOC is a modular {\tt FORTRAN77} program using a random number
generator described in \citet{random}.  The program produces an ASCII
output containing one block of data for each synthetic ``event''. One
event corresponds to one SN. Each SN is characterized by its redshift
and SN type (i.e whether is a Type Ia, Ibc, IIn, IIP, IIL or 'SN1987A
like').  The luminosity distance corresponding to the SN redshift is
calculated from the user supplied set of cosmological parameters where
we assume a standard Friedmann-Lema\^{\i}tre (FL) universe with
arbitrary choice of ``dark energy''. In order to estimate effects
along the line-of-sight, e.g., from gravitational lensing and dust
extinction, the light~beam is traced backwards in time from the
observer to the host galaxy, taking into account the non-homogeneity
of the universe and possible intervening matter. This procedure is
shown schematically in Fig.~\ref{fig:schematic}.

The program sequence is as follows:

\begin{itemize}
\item Initialization
\item Read user-supplied parameters (see next section)
\item Loop: event generator
\begin{enumerate}
 \item Select SN type
 \item Generate SN redshift, $z$
 \item Calculate distance and lookback time to $z$
 \item Calculate unperturbed SN lightcurve in a ``redshifted'' B-band filter\footnote{Transmission function satisfying
B$^z(\lambda)$=B$({\lambda \over 1 + z})$}
 \item Perform ray-tracing
 \begin{enumerate} 
  \item Calculate gravitational lensing magnification
  \item Perform differential dust extinction calculation
  \item Estimate dimming due to exotics, e.g., photon-axion oscillations
 \end{enumerate}
 \item Calculate apparent SN brightness in BVRIJ filters
 \item Generate host galaxy properties
\end{enumerate}
\end{itemize}


\section{The user supplied parameters}
Table \ref{tab:snoc_input} in Appendix A summarizes the user supplied parameters for
the simulation. These include the following main items:
\begin{itemize}
\item The number and type of SNe to be generated. The SN types supported
      are Ia, Ibc, IIn, IIP, IIL and core collapse SNe resembling SN 1987a.

\item The redshift distribution to be used. The options include:
\begin{enumerate} 

\item Uniform distribution within a chosen range
\item Tabulated redshift distribution \item Listed individual
redshifts 
\item Gaussian probability distribution function
 with given mean value and standard
deviation 
\item Redshift distribution generated from star formation
rates [Ref.\,~\cite{DahlenFransson}] \end{enumerate}
\item Options for generating the SN lightcurve
\item Cosmological parameters: $[H_0,\Omega_M,\Omega_X,\alpha_X(z)]$
\item Gravitational lensing. The user supplies the relative fraction 
of galaxy halos parametrized  as singular isothermal spheres (SIS), 
Navarro-Frenk-White profiles (NFW), compact objects  or spheres with 
uniform mass density.
\item Dust extinction parameters: mean free paths and frequency dependence
         for host galaxy dust extinction, dust in 
       intervening galaxies and in the intergalactic medium.
\item Choice of observed filters (K-corrections)
\item Galaxy parameters: luminosity and mass distribution
\item SN search conditions
\end{itemize}

\section{The event generator}
\subsection{Supernova Type and redshift distribution}

The SN type is selected as either individual type (e.g., 'Ia') or
a group of SNe (e.g., 'core') which generates core collapse SNe
with a differential rate as in \citet{DahlenFransson} and 
\citet{Dahlen&Goobar}.

Several redshift distributions are provided by the code: uniform or Gaussian
distributions as well as physically motivated redshift distributions according to:

\begin{equation}
{dN \over dz} = {4 \pi c d_L^2(z) \over H(z) (1+z)^2}\cdot {r_{\rm SN}(z) \over 1+z},
\end{equation}

where $r_{\rm SN}$ is the rate of SNe in comoving coordinates. In SNOC we
implemented the 
models of the evolution of the comoving rate of SN explosions in
\citet{DahlenFransson}. In these models an additional user supplied 
parameter $\tau$ indicates the minimum time needed to form the binary
systems believed to be the progenitors of Type Ia SN
explosions. Typical values range from 0.3 to 3.0 billion years.

It is also possible to generate SNe with redshifts from a user 
provided list of redshifts or user defined distributions.

\subsection{The luminosity distance and apparent magnitude: cosmological models}
The luminosity distance and the lookback time to the event are
calculated assuming an isotropic and homogeneous universe. In such a
FL universe, the scale factor of the
universe and thereby the distance estimates are functions of the
Hubble constant $H_0$ and the energy components of the universe 
\citep[see, e.g.,][]{BergstromGoobar1999}.

The energy density components are treated as perfect fluids
characterized by their equation of state, $p_i=\alpha_i\rho_i$. Besides
the mass density $\Omega_M$ ($\alpha_M=0$), a second fluid is used
in the simulation, $\Omega_X$, with a user specified parameter
$\alpha_X(z)$.  Thus, SNOC can generate luminosity distances for any
choice of $\Omega_M$ and $\Omega_X$ including the cosmological
constant ($\alpha_X=-1$) or a time dependent equation of state parameter
as predicted by {\em quintessence}. In addition to the FL luminosity
distance, or {\em filled beam} distance, the distance in absence of
any focusing of the light~beam, the {\em empty beam} distance, is
calculated.

The apparent magnitude $m$ of a SN at redshift $z$, assuming the
cosmology $\vec{\theta}=[\Omega_M,\Omega_X,\alpha_X(z)]$, is given by
\begin{eqnarray}
    m(\vec{\theta},{\cal M},z)&=&
    {\cal M}+5\log_{10}\left[d'_L(\vec{\theta},z)\right],\\
    {\cal M}&=&25+M+5\log_{10}(c/H_0),\label{eq:Mscript}
\end{eqnarray}
where $M$ is the absolute magnitude of the SN, and 
$d'_L\equiv H_0\,d_L$ is the \emph{$H_0$-independent} luminosity distance, 
where $H_0$ is the Hubble parameter.\footnote{In the expression for 
${\cal M}$, the units of $c$ and $H_0$ are km s$^{-1}$ and 
km s$^{-1}$ Mpc$^{-1}$, respectively.} 
Hence, the intercept ${\cal M}$ contains the ``unwanted''
parameters $M$ and $H_0$ that apply equally to all magnitude
measurements [we do not consider evolutionary effects $M=M(z)$].
The $H_0$-independent luminosity distance $d'_L$ is given by
\begin{eqnarray}\label{eq:lumdist}
    d'_L&=&\left\{
    \begin{array}{ll}
      (1+z)\frac{1}{\sqrt{-\Omega_k}}\sin(\sqrt{-\Omega_k}\,I) , &
      \Omega_k<0\\
      (1+z)\,I , & \Omega_k=0\\
      (1+z)\frac{1}{\sqrt{\Omega_k}}\sinh(\sqrt{\Omega_k}\,I) , &
      \Omega_k>0\\
    \end{array}
    \right. \\
    \Omega_k&=&1-\Omega_M-\Omega_X,\\
    I&=&\int_0^z\,\frac{dz'}{H'(z')} ,\\
    H'(z)&=&H(z)/H_0=\nonumber\\
    &&\sqrt{(1+z)^3\,\Omega_{M}+
    f(z)\,\Omega_{X}+(1+z)^2\,\Omega_{k}}, \\
    f(z)&=&\exp\left[3\int_0^z\,dz'\,\frac{1+\alpha_X(z')}{1+z'}\right] ,
\end{eqnarray}
where we consider an equation of state with arbitrary $z$-dependence.

\subsection{Ray-tracing}
The ray-tracing is discussed in detail in
Sect.~\ref{section:ray-trace}. Here, we summarize its key elements. The
light-beam path is divided into 'cells' corresponding approximately to
average distances between galaxies. Each cell is characterized by
\begin{enumerate}
 \item The mass density profile, the source of gravitational lensing
 \item The dust density and differential extinction coefficient,
 treated separately for the actual galaxy in the cell center and the
 region around it 
 \item The average magnetic field and electron
 density, relevant in exotic scenarios such as the photon-axion
 oscillation mechanism discussed in Sect.~\ref{section:axion}
\end{enumerate}
\subsection{Supernova and Galaxy apparent magnitudes: K-corrections}
The SN and host galaxy apparent magnitudes through a standard (or user
specified) broad-band filter is calculated using cross-filter
K-corrections as described in \citet{KimGoobarPerlmutter96}.  Thus,
the apparent magnitude of a SN observed at a redshift $z$ through a
filter Y defined through the filter function $S_{Y}(\lambda)$ becomes:
\begin{equation}
    m_Y(z) = 
    {\cal M}_X +5\log_{10} \left[d'_L(\vec{\theta},z) \right] + K_{XY},
\end{equation}
where the restframe absolute magnitude of the SN is measured in the
filter X with a filter transmission function $S_X(\lambda)$.

Assuming a spectral template for each SN type, $F_{\rm SN}$ and 
an arbitrary differential attenuation $0 < \epsilon_i(\lambda) < 1$ the 
K-correction becomes:
\begin{eqnarray*}
K_{XY}^i & = &  -2.5 \log_{10} \left(
{\int{ {\cal V}(\lambda) S_X (\lambda) \lambda d \lambda} \over
 \int{ {\cal V}(\lambda) S_Y (\lambda) \lambda d \lambda}  } \right) \nonumber \\
& &  + 2.5 \log_{10} \left(
{\int{ F_{\rm SN}(\lambda) S_X (\lambda) \lambda d \lambda} \over
 \int{ F_{\rm SN}(\lambda) \epsilon_i(\lambda) S_Y (\lambda(1+z)) \lambda d \lambda} } 
\right),
\end{eqnarray*}
where the photometry system is normalized in the Vega system, ${\cal V}$.

By default, SNOC calculates K-corrections for K$_{\rm BY}^i$, where Y
includes the standard broad-band filters B, V, R, I and J. In order to
assess the color corrections from the various extinction sources we
list the results separately:
\begin{itemize}
\item[0)] K$_{\rm BY}^0$: no extinction ($\epsilon_0=1$)
\item[1)] K$_{\rm BY}^1$: host galaxy extinction
\item[2)] K$_{\rm BY}^2$: extinction from galaxies along the line-of-sight
\item[3)] K$_{\rm BY}^3$: extinction from intergalactic dust
\item[4)] K$_{\rm BY}^4$: differential effects from photon-axion oscillations
\end{itemize}

Finally,  K$_{\rm BY}$ (no index) shows the combined effect on the observed
magnitudes from all the considered attenuation sources. 

Similarly, representative galaxy templates for E, Sbc, Scd and Im
galaxies from \citet{col80} are used to calculate the observed
colors of the SN host galaxies.


SNOC generates simulated lightcurve data with a user specified time
sampling, date of discovery and noise contamination. For Type Ia SNe,
K-corrections for BVRIJ may be calculated starting from up to 19 days
prior to maximum up to 70 days past maximum in the restframe system
using a spectral template from \citet{nugent}.\footnote{The spectral
template of day 70 is used for later epochs.} For other types of SNe,
approximate K-corrections are calculated from a set of (truncated)
black-body functions, following the approach in
\citet{DahlenFransson}.

\subsection{Simulation of supernova search conditions}
Two samplings of the SN
lightcurve are generated randomly using a user specified time gap. 
SNOC can thus be used to optimize the search strategy for finding SNe
at any given redshift. The tuning parameters are the broad band filter chosen, the limiting magnitude of the 
search and the time between the reference and the discovery images. The method can also be used
to study the contamination from core collapse SNe in the discovery sample 
\citep{Dahlen&Goobar}.


\section{The elements of the ray-tracing section}\label{section:ray-trace}
\subsection{Gravitational lensing}
When simulating observations of distant sources, it is necessary to 
take into account the fact that the Universe is inhomogeneous. First, 
the inhomogeneities might influence the overall expansion rate of the 
Universe, the so called {\em averaging problem}. Second, 
gravitational lensing effects due to the inhomogeneities will cause 
a dispersion in the observed source luminosities.

In SNOC we employ a method proposed by  
\citet{art:HolzWald1998} for examining lensing effects. In this 
method, the overall expansion is assumed to follow that of the 
average background geometry. That is, in the following we will 
neglect the averaging problem and only consider effects from gravitational 
lensing.\footnote{For general references on gravitational lensing see, 
e.g., the textbook by \citet{book:Schneider}.}

\subsubsection{The Method} 
The method of Holz \& Wald can be summarized as follows: First, a
FL background geometry is
selected, i.e., we specify the space-averaged energy density content of the 
Universe, $\Omega_M$ and $\Omega_X$.
Inhomogeneities are accounted for by specifying matter
distributions in spherical cells which have an average energy density equal to
that of the underlying FL model. Thereafter, an infinitesimal beam focused 
at the observer is integrated backwards using the geodesic deviation
equation by being sent through a series of cells, each time
with a randomly selected impact parameter with respect to the
matter distribution in the cell. 
Finally, the resulting area, $A$, at the source redshift is considered.

In a homogeneous FL universe, the ray bundle will have an area
corresponding to the filled beam area, $A_{\rm FL}$.
Since the apparent luminosity of a source
is proportional to the inverse of
the area of the beam, we define the magnification as  
\begin{equation}
    \mu =\frac{A_{\rm FL}}{A}.
\end{equation}
By using Monte-Carlo techniques to trace a
large number of light~rays, statistics for the apparent
luminosity of an ensemble of sources is obtained. 

As noted in Holz \& Wald, some care is needed when interpreting
results as probability distributions. This is due to the fact that
individual ray bundles do not correspond to random source positions.
Since the fraction of the source sphere being sampled by a beam is
proportional to the area of the beam, i.e., inversely proportional to
the magnification, magnified sources will be over-represented if not
compensated for. This is done by designating a probability
proportional to the area of the ray bundle to keep each event. Further
discussions of the statistical weight given to random lines-of-sight
can be found in \citet{EH86} and \citet{bergstrom2000}.

In order to validate our implementation of the method, we have
performed a number of consistency tests. The interested reader is referred to
\citet{bergstrom2000}.

\subsubsection{Matter distributions}

The matter distribution in the cells can be chosen to be in arbitrary
fractions 
of point-masses, uniform spheres, singular isothermal spheres (SIS) or the 
Navarro-Frenk-White (NFW) density profile \citep{art:NFW}. 
Matter distribution parameters such as the scale radius
of the NFW halo and the cut-off radius of the SIS halo are determined
from distributions reflecting real galaxy properties.

One of the advantages of the method of Holz \& Wald is that any mass distribution
and number density, including possible redshift dependencies,
can easily be implemented and used as long as the 
average energy density in each cell agrees with the underlying
FL model.

We derive a galaxy mass distribution, $dn/dM$, 
by combining the Schechter luminosity function 
\citep[see, e.g.,][Eq.~5.129]{book:Peebles} 
with a mass-to-luminosity ratio
\begin{equation}
  M\propto L^{\frac{1}{1-\beta}}.
\end{equation}
For elliptical galaxies in the fundamental plane, we can, e.g., choose
$\beta = 0.25$ \citep[see][Eq.~3.39]{book:Peebles}. The normalization
of the resulting galaxy mass distribution, i.e., the characteristic
galaxy mass, $M_*$, is determined from observational constraints on
the galaxy number density.

\subsubsection{Foreground cluster}

Since all cells have the same average density as the underlying FL model, the 
average luminosity will be equal to the one obtained in the homogeneous case.  
To simulate the increased gravitational lensing effects due to a specific matter 
distribution along a given line-of-sight,
the user can specify the properties of a cell at a given redshift 
to mimic a foreground galaxy or cluster.
The matter distribution is in the form of an isothermal sphere,
\begin{equation}
  \rho_{{\rm SIS}}(r)=\frac{v^{2}}{2\pi }\frac{1}{r^{2}},
\end{equation}
where $v$ is the line-of-sight velocity dispersion of the mass particles.
Thus, the velocity dispersion sets the concentration of the cluster, the mass
sets the cut-off radius and the radius of the cell sets the size of the
image field.

\subsubsection{Multiple images}

Since we in principle only are able to follow infinitesimal light-rays in 
SNOC, we have to use some approximations when trying to get information on 
multiple image systems. The main approximation is that we assume that in
cases of strong lensing, the effects from one close encounter is dominant,
i.e., the one-lens approximation. With this simplification, we can use the 
information from the magnification to derive quantities for systems with 
finite separations. In order to do this, we need to be able to derive analytical 
relations between different observables, e.g., the magnification and the image separation. 
Here we show how this is done for the case of SIS lenses.

When running SNOC, we want to answer the following questions:
\begin{itemize}
\item Will the particular event correspond to the primary image in a multiple 
lens-system? If this is the case:
\item What is the flux of the second image?
\item What is the separation between images?
\item What is the time-delay between images?
\end{itemize}
Employing the one-lens approximation with a SIS lens, we will
have multiple imaging whenever $\mu_1>2$ and we can relate the
magnification of the secondary image, $\mu_{2}$, to that of the first by
\begin{equation}
    |\mu_{2}|=\mu_{1}-2.
    \label{eq:sismu2}
\end{equation}
Furthermore, the image separation is given by
\begin{equation}
  \label{eq:delthe}
  \Delta\theta =8\pi\left(\frac{v}{c}\right) ^2\frac{D_{ds}}{D_s},
\end{equation}
where $D_s$ and $D_d$ are the angular size distances, to 
the source ($z_s$), the lens ($z_d$), while $D_{ds}$ represents
the angular size distance between the lens and the source.
The time delay for the two images is
\begin{equation}
  \Delta t=\left [4\pi\left (\frac{v}{c}\right )^2\right ]^2
            \frac{D_dD_{ds}}{D_s}(1+z_d)\frac{2}{\mu_1-1}.
\end{equation}
Therefore, in order to compute the quantities of interest, we need to
pick a lens redshift and velocity dispersion from some reasonable
distributions for every case where $\mu_1>2$. The distributions
will in general be functions of the cosmology, the mass distribution
of the lenses and the source redshift.
The differential probability for multiple imaging is in the general case
given by
\begin{equation}
  dP\propto\sigma (z_d,z_s)\frac{dn}{dM}dM (1+z_d)^3 dV,
\end{equation}
where $\sigma (z_d,z_s)$ is the cross-section for multiple imaging and
$n$ is the comoving number density of lenses, assumed to be
constant. For SIS lenses, the cross-section is given by
\begin{equation}
  \sigma (z_d,z_s)\propto\left(\frac{v}{c}\right)^4D_{ds}^2.
\end{equation}
Since $dV\propto D_d^2\frac{dt}{dz_d}dz_d$ and $n$ is independent of lens
redshift we can use
\begin{equation}
  dP(z_d)\propto D_{ds}^2D_d^2(1+z_d)^3\frac{dt}{dz_d}dz_d,
\end{equation}
as our probability distribution for $z_d$. The probability distribution for $v$
is given by
\begin{equation}
  dP(v)\propto \frac{dn}{dM}\frac{dM}{dv}\left(\frac{v}{c}\right)^4dv,
\end{equation}
where the velocity dispersion is related to mass by
\begin{equation}
  \frac{v}{v_*}=M^{\gamma (1-\beta)},
\end{equation}
\citep[see][Sect.7.1]{bergstrom2000}.

%
%

\subsection{Extinction by dust along the line-of-sight}

An important effect to consider when observing SNe is the absorption
of light due to dust. With SNOC, it is possible to take the effects
from three different extragalactic dust components into account:
\begin{enumerate}
\item A homogeneous intergalactic component.
\item A host galaxy dust component.
\item Dust in galaxies between the source and the observer. 
\end{enumerate}
Properties such as the exponential extinction scale-length and the 
frequency dependence can be set
separately for each component. The frequency dependence is computed using the 
parametrization in \citet{Cardelli}.

\subsubsection{Grey dust}

For a  given emission redshift $z_s$, the attenuation $\Delta m_d$ at observed
wavelength $\lambda_o$ 
due to dust can be written 
\begin{eqnarray}
\ & \Delta m_d(z_s,\lambda_o) = \ \ \ \ &  \nonumber  \\
& -2.5\log_{10}\left[e^{-C\int_0^{z_s}\rho_{\rm dust}(z)a(\lambda_o/(1+z),R_V)
h(z)dz}\right], 
\label{eq:deltam}
\end{eqnarray}
where $\rho_{\rm dust}(z)$ is the physical dust density at redshift $z$, $a(\lambda,R_V)$
is the wavelength-dependent attenuation, and the cosmology-dependent function 
$h(z)$ is given by
\begin{equation}
h(z)=[H(z)\left(1+z\right)]^{-1}.
\end{equation}
The reddening parameter $R_V$ is defined by
\begin{equation}
  \label{eq:rv}
  A_V=R_VE(B-V),
\end{equation}
where $A_V$ is the V-band extinction and
\begin{equation}
  E(B-V)=(B-V)-(B-V)_i 
\end{equation}
with $(B-V)_i$ being the intrinsic (unobscured) color.

The normalization constant $C$ is related to the overall magnitude of the 
extinction, i.e., the exponential extinction scale-length.\footnote{ 
This can be fixed, e.g., by demanding that a given cosmology reproduces the observed
luminosity distance at $z\sim 0.5$, i.e., that dust extinction can 
explain the dimming of the presently observed SN sample, 
otherwise attributed to the ``concordance'' cosmology
$\Omega_M=0.3$, $\Omega_\Lambda=0.7$.}

In a manner similar to that used in computing gravitational lensing effects,
the integral in Eq.~(\ref{eq:deltam}) is performed
numerically by following individual light paths through a large
number of cells containing galaxies and intergalactic dust. Through
each cell the background cosmology, the wavelength of the photon and the
dust density are updated, and the contributions from each cell added. 
Note that the model is approximately valid also for a patchy dust distribution, 
as long as the scale of 
inhomogeneities is small enough, i.e., $1/\sqrt{N}\ll 1$ where $N$ is the 
number of dust clouds intersected by the light-ray. 

The intergalactic dust density is parametrized by
$\rho_{\rm dust} = \rho^\circ_{\rm dust}(1+z)^{f(z)}$, where 
\begin{eqnarray}\label{eq:dustpara}
 f(z) =  
      \left\{ 
        \begin{array}{lc}
          \alpha ={\rm const. \ for}&  z<z_{\rm lim} \\
          \beta ={\rm const. \ for}&  z>z_{\rm lim}
       \end{array}
        \right.   
\end{eqnarray}
$z_{\rm lim}$ is a user defined redshift where the intergalactic
dust distribution is assumed to have changed properties.
It is also possible to supply a user specified set of values of the 
dust density $\rho_{\rm dust}(z)$ and the frequency dependence as a
function of redshift. 
%


\subsubsection{Host galaxy dust}

The absorption in the host galaxy will be different for different SN
types. This is because core collapse SNe (Type II, Ibc etc.) occur in
late type, star forming, dusty galaxies whereas Type Ia can occur in
any type of host galaxy. In the current version of SNOC, we have used
Fig.~6 in \citet{matteucci} to make a first rough estimate of the
probability for a Type Ia SNe (up to $z<4$) to occur in an early type
galaxy as
\begin{equation}
  \label{p}
  p_E(z)=p_E(0)+q_1\cdot z  
\end{equation}
where $p_E(0)=0.5$ and $q_1=0.125$. This simplistic approach could
without doubt be improved on in the future.

Also, the amount of dust and the SN rate depend on the size of the
galaxy (i.e., the star content).  Using Monte-Carlo methods, the
galaxy type, dust content, SN position and disk inclination are
determined and the absorption is calculated.

We model the dust distribution in late type galaxies as a double exponential disk
\begin{equation}
  \label{eq:dustdist}
  \rho_{\rm dust}=\rho^0_{\rm dust}e^{-r/r_0-|\zeta|/\zeta_0},
\end{equation}
where $r$ and $\zeta$ are cylindrical coordinates. We set 
$\zeta_0=0.1$ kpc. In order to determine $r_0$, we assume that the
amount of dust is proportional to the amount of stars in the
galaxy. If stars are distributed as
\begin{equation}
  \label{eq:stardist}
  n_*=n_{*}^0e^{-r/r_0-|\zeta|/\zeta_0},
\end{equation}
the luminosity, $L$, will be proportional to $\int n_* dV\propto
r_0^2\cdot \zeta_0$. If $\zeta_0$ is fixed, we have $r_0\propto\sqrt{L}$.
That is, we set $r_0=\sqrt{L}\,r_{*}^0$ where $r_{*}^0=5$ kpc is the
scale radius for a $L=L_*$ galaxy.  We determine $\rho^0_{\rm dust}$
by setting the exponential scale-length at the galaxy center. The disk
is truncated at $r_{\rm max}=20$ kpc and $\zeta_{\rm max}=3$ kpc.

We let the probability for a SN to occur in a specific galaxy be
proportional to the galaxy luminosity in order to allow for SNe to be
more common in star-rich, luminous galaxies. Core collapse only occur
in the disk, closely following the dust distribution in
Eq.~(\ref{eq:dustdist}), but we do not allow any SNe inside
$r=0.3$ kpc.  We let a fraction (87.5\,\%) of Type Ia SNe follow the
dust distribution in Eq.~(\ref{eq:dustdist}), only that we set
$\zeta_0=0.35$ kpc, and the remaining fraction (12.5\,\%) occur in a bulge
according to the probability distribution
\begin{equation}
  \label{eq:bulge}
  p(R)\propto\frac{1}{R^3+0.7^3},
\end{equation}
where $R$ is the radial coordinate in spherical coordinates (expressed
in kpc). The bulge is truncated at $R_{\rm max}=3$ kpc. For further
reference regarding dust and SNe distribution in host galaxies, see
\citet{Hatano}.
After determining the dust distribution and SN position in the galaxy,
we assume a random inclination of the disk and integrate the
absorption along the path of the light-ray through the disk.


\subsubsection{Dust in intervening galaxies}

In each cell, the galaxy type is selected using a Monte-Carlo method. 
We use a parametrization where the fraction of early type, dust-free galaxies 
is given by $f_E(z)=f_E(0)-q_2*z$, where $f_E(0)=0.3$ and $q_2=0.05$. If the galaxy 
is a spiral, the absorption due to the dusty disk is computed (if the light-ray passes
it).  
Equivalent to the case of host galaxies, we model the dust distribution in 
intervening spiral galaxies by a double exponential disk with random
inclination, see Eq.~(\ref{eq:dustdist}).

\subsection{Photon-axion oscillations}\label{section:axion}

The general structure of SNOC, i.e., the tracing of light through
successive cells, makes it relatively easy to add more refinement to
the propagation code. As an example we mention here an exotic,
hypothetical process that could affect light propagation on
cosmological scales.  

It has recently been proposed that the observed faintness of
high-redshift SNe could be attributed to the mixing of photons with a
light axion in an intergalactic magnetic field
\citep{csaki}. We compute the mixing probability using the formalism of density
matrices
\citep[see, e.g.,][]{sakurai}.  
We define the mixing matrix as 
\begin{equation}
  \label{eq:M} 
  Q=\left(\begin{array}{ccc}
       \Delta_{\perp} & 0 & \Delta_{\rm M}\cos\alpha\\
       0 & \Delta_{\parallel} & \Delta_{\rm M}\sin\alpha\\
       \Delta_{\rm M}\cos\alpha & \Delta_{\rm M}\sin\alpha & \Delta_{\rm m}
       \end{array}\right).
\end{equation}
The different quantities appearing in this matrix are given by 
\begin{eqnarray}
  \label{eq:terms} 
  \Delta_{\perp} & = & -3.6\times 10^{-25}\left(\frac{\omega}{1\,{\rm eV}}\right)^{-1}
\left(\frac{n_{\rm e}}{10^{-8}\,{\rm cm}^{-3}}\right){\rm cm}^{-1},\\
\Delta_{\parallel} & = & \Delta_{\perp},\\
\Delta_{\rm M} & = & 2\times 10^{-26}\left(\frac{B_{0,\perp}}{10^{-9}\,{\rm G}}\right)
\left(\frac{M_{\rm a}}{10^{11}\,{\rm GeV}}\right)^{-1}{\rm cm}^{-1},\\
\Delta_{\rm m} & = & -2.5\times 10^{-28}
\left(\frac{m_{\rm a}}{10^{-16}\,{\rm eV}}\right)^2
\left(\frac{\omega}{1\,{\rm eV}}\right)^{-1}{\rm cm}^{-1},
\end{eqnarray}
where $B_{0,\perp}$ is the strength of the magnetic field perpendicular to the
direction of the photon, $M_{\rm a}$ is the inverse coupling between the 
photon and the axion, $n_{\rm e}$ is the electron density,
$m_{\rm a}$ is the axion mass and $\omega$ is the
energy of the photon.
The angle $\alpha$ is the angle between the (projected) magnetic field
and the (arbitrary, but fixed) perpendicular polarization vector.
The time-evolution of the density matrix $\rho$ is 
given by
\begin{equation}
  \label{eq:rhoeq}
  {\rm i}\delta_t\rho =\frac{1}{2\omega}[Q,\rho], 
\end{equation}
with initial conditions
\begin{equation}
  \label{eq:rho0} 
  \rho_0=\left(\begin{array}{ccc}
       \frac{1}{2} & 0 & 0\\
       0 & \frac{1}{2} & 0\\
       0 & 0 & 0
       \end{array}\right).
\end{equation}
Here the three diagonal elements refer to two different polarization
intensities and the axion intensity respectively.
In each cell, we solve the 9 coupled (complex) differential equations and
update $\rho$ and $Q$. This method ensures that the full frequency dependence 
of the effect is taken into account, as shown in \citet{axion}.  

\newcommand{\om}{\Omega_M}
\newcommand{\ox}{\Omega_X}
\newcommand{\sm}{{\cal M}}
\newcommand{\negml}{{\cal L}}
\newcommand{\ls}{d_L'}

\section{Fitting cosmological parameters}
The SNOC package also includes a software tool for cosmology fitting,
based on the output files from the Monte-Carlo program. In its
current state two different types of maximum-likelihood analysis can
be done using different measurement variables.

\paragraph{Type Ia supernovae as standard candles}

The analysis tool can be used to fit the cosmological parameters
${\cal M}$, $\Omega_M$, $\Omega_X$, $\alpha_X$, and parameters related to
the quintessence models presented in \citet{GoliathEtAl:2001} and
\citet{Eriksson:2002}, by using simulated Type~Ia events where the
magnitude $m_i$ (in the redshifted B-band)  and the redshift $z_i$ are
considered to be the measurement variables for each event $i$. The
negative log-likelihood function will in this case take the following
appearance
\begin{equation}
  \begin{split}
    \negml  & =
    -N\ln\left(\frac{1}{\sqrt{2\pi}\sigma}\right) +\\
    & + \frac{1}{2\sigma^2}
    \sum^N_{i=1}\left[m_i - 5\log_{10}\ls(\om,\ox,\alpha;z_i) -
      \sm\right]^2\, ,
  \end{split}
  \label{eq:negml}
\end{equation}
where $N$ is the number of events, $\sigma$ the statistical
uncertainty for each event, $m_i$ and $z_i$ and finally the luminosity
distance, $\ls$, is given by Eq.~\eqref{eq:lumdist}.

It is also possible to use a different, more general, likelihood
function that also takes into account the fact that the data could be
subject to systematic effects from gravitational lensing. 
This procedure is further described in \citet{Maglensed:2002}.

\paragraph{Cosmological parameters from lensed supernovae}

The other way to use the analysis tool is to use lensed SNe as
is described in detail in \citet{Goobar:2002}. In this case
core-collapse SNe as well as Type Ia's can be used, and in its
present state the software allows fitting of the parameters $H_0$,
$\om$, $\ox$ and $\alpha_X$(=constant), by using multiply imaged
high-$z$ SNe in spherically symmetric lensing systems producing
only two or ring-like images. The observables are assumed to be the
source redshift, the redshift of the lensing galaxy, the image
separation, $\Delta\theta$, the time delay, $\Delta t$, and the
flux-ratio, $r_f$, between the images.

The negative log-likelihood function for this analysis can be
expressed as
\begin{equation}
  \negml = \sum^N_{i=1}\frac{(R_{\mathrm{exp}}^i -
    {\cal R}^i)^2}{2\sigma^2}\, ,
\end{equation}
where $R_{\mathrm{exp}}^i$ and ${\cal R}^i$ can be written as
\begin{eqnarray}
  R_{\mathrm{exp}}^i & = & 2 {\Delta t \over {\Delta \theta}^2}\left({1 +
      z_d \over f(r_f)}\right)^{-1}\quad{\rm and} \\
  {\cal R}^i & = & {D_{d}D_{s}  \over D_{ds}}. 
\end{eqnarray}
The function $f$ depends on the flux-ratio and the applied lens model.

\subsection{Minimizing the negative log-likelihood function}
The user may choose between the 
three implemented methods to minimize the
negative log-likelihood function, independently of what it looks like.
\begin{enumerate}
  \item A simple method to minimize the negative log-likelihood
    function is to calculate its value in a grid, specified by start and
    end values for each parameter and the number of grid steps in each
    dimension. In the output file (see Appendix~B) the
    result array is saved as an ASCII vector.
  \item The basic idea of the \emph{Davidon Variance
    Algorithm} \citep{Davidon:1968} is to calculate the covariance
    matrix by an iterative algorithm. The matrix is obtained by using
    only the function values and the gradient, and the minimum value
    is calculated simultaneously as the algorithm converges. This
    method is much faster in finding the minimum than the grid search,
    the disadvantage however, is that a parabolic shape is assumed for
    the negative log-likelihood function and the estimated values for
    the covariance matrix are somewhat misleading if this is not the
    case.
  \item \emph{Powell's Quadratically Convergent
      Method} \citep{Powell:1964} is a multidimensional minimization
    algorithm that does not require any derivatives. Basically it is
    a method to choose in an efficient way a direction in the
    minimization space and then use linear minimization in that
    direction. This method will however not calculate any covariance
    matrix, only the minimum point of the function.
\end{enumerate}

\section{Examples of use}
As an example of the usage of SNOC, we show how to create a data
sample of 2000 Type Ia SNe with a uniform distribution $0.1<z<2$ and
fit the parameters $\Omega_M$ and $\alpha_X$ (assuming a flat
geometry) by performing a maximum likelihood analysis.

In the SNOC input file (see Appendix A), we set: 
\begin{center}
\begin{tabular}{lll} 
{\tt NUMSNE} & 2000 &\\ 
{\tt GENDIS} & 'uni' &\\ 
{\tt MAXZ} & 2.0 &\\
{\tt MINZ} & 0.1 &\\
{\tt SNTYPE} & 'Ia' &\\ 
{\tt INTSIG} & -1.0 & (default intrinsic magnitude spread)\\
{\tt FILTER} & 'BZ' & (redshifted B filter)\\  
{\tt HUBBLE} & 0.65 &\\
{\tt OMEGAM} & 0.3 &\\
{\tt OMEGAX} & 0.7 &\\
{\tt EOSTYPE} & 'con' &\\
{\tt ALPHAX0} & -1.0 & (cosmological constant)\\                  
{\tt REJTYPE} & 3 & (keep only primary images)\\
{\tt DOFRAC} & 1 & (fractional distribution of lenstypes)\\           
{\tt FRACPOI} & 0.2 & (20\,\% point-masses)\\
{\tt FRACNFW} & 0.8 & (80\,\% NFW halos)\\
\end{tabular}
\end{center}
All other logical options should be set to 'false' or '0'. We have
used a ``standard'' cosmology with $\Omega_M=0.3$ and a pure
cosmological constant $\Omega_\Lambda =0.7$ and computed lensing
effects with 20\,\% point-masses and 80\,\% NFW halos. 

\begin{figure}
\includegraphics[width=\hsize]{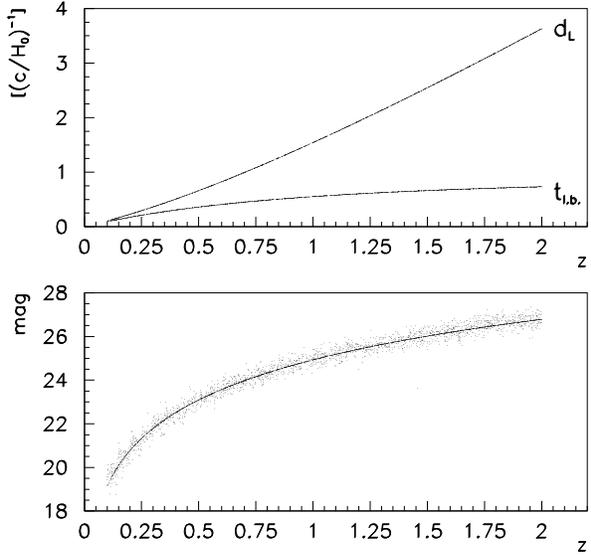}
\caption{In the upper panel the luminosity distance,
$d_L$, and the lookback time $t_{\rm l.b.}$ (in units of $c/H_0$) 
are given as a function of redshift. In the lower panel the theoretical
magnitude-redshift relation for Type Ia SNe (full line) is plotted together 
with the magnitude of each individual SNe including the intrinsic
dispersion and gravitational lensing effects.}
\label{fig:hubble}
\end{figure}

In Fig.~\ref{fig:hubble}, we show the Hubble diagram obtained by
plotting the various quantities of interest from the SNOC output ASCII
file (see Appendix B). In the upper panel, the luminosity distance,
$d_L$, and the lookback time $t_{\rm l.b.}$ (in units of $c/H_0$) are
given as a function of redshift by plotting {\tt dL} and {\tt lookb}
vs {\tt zs}.  

In the lower panel, the full line is the theoretical
magnitude-redshift relation for a typical Type Ia SNe observed at
maximum luminosity in the restframe B band. This is obtained by
plotting {\tt BZmag}-{\tt intsig} vs {\tt zs}. In the same panel the
magnitude of each individual SNe is shown, including the intrinsic
dispersion and gravitational lensing effects, i.e., we have plotted
{\tt BZmag}+{\tt linsdm} vs {\tt zs}.

We can now use the SNOC analysis tools to fit, e.g., 
$\Omega_M$ and $\alpha_X$, assuming $\Omega_M+\Omega_X =1$.

In the input for the analysis tool (Appendix A), we set: 
\begin{center}
\begin{tabular}{lll} 
{\tt nrsne} & 2000 &\\
{\tt nrparsigma} & 2 &\\
{\tt flat} & (assuming $\Omega_M+\Omega_X =1$)\\
{\tt w\_0} & -1.0 0.0 200 -1 &\\
{\tt omega\_M} & 0.0 0.4 200 0.3 &\\
{\tt NUISM} & &\\ 
{\tt nointsig} & &\\
{\tt gauss} & &\\
{\tt nodavidonmin} & &\\ 
{\tt gridsearch} & &\\
{\tt nopowellmin} & &\\ 
\end{tabular}
\end{center}

\begin{figure}
\includegraphics[width=\hsize]{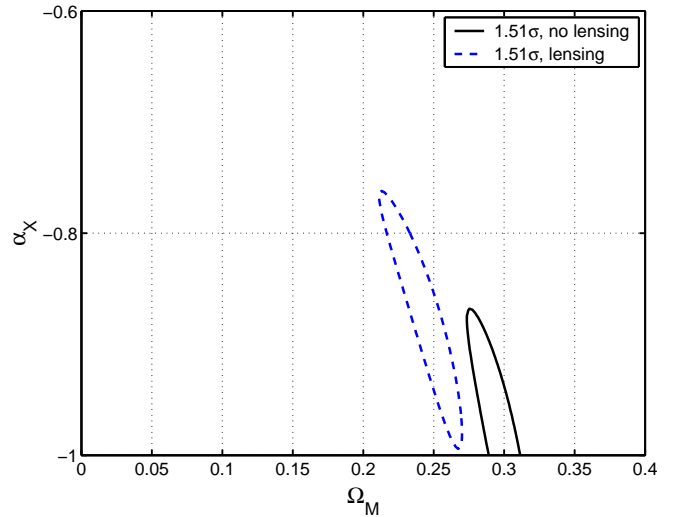}
\caption{Three-parameter fits in the $(\Omega_M,\alpha_x)$-plane using
  the data set shown in Fig.~\ref{fig:hubble}. The contours show the
  1.51$\sigma$ level, which corresponds to a probability content of
  $\sim68\,\%$ of including both parameter in the region.}
\label{fig:omwfit}
\end{figure}

Fig.~\ref{fig:omwfit} shows the confidence contours produced from
this three-parameter fit, where $\sm$ has been treated as a nuisance
parameter. The solid contour that is centred around the input values
$\om=0.3$, and $\alpha_X=-1.0$ is the fitted confidence region when
the lensing dispersion is not considered, i.e., the keyword
\texttt{nolenseffect} has been specified in the input file. The dashed
contour shows instead the fitted result when the
\texttt{addlenseffect} parameter has been specified in the input file
and the lensing effects are considered. As can be seen, this will
induce a significant bias in the parameter determination, an effect
discussed in further detail in~\citet{Maglensed:2002}. Both
contours shows the
\begin{displaymath}
  \textrm{min}(\negml) + \frac{1.51^2}{2}\,
\end{displaymath}
level of the negative log-likelihood surface, $\negml$, defined in
Eq.~\eqref{eq:negml}. This level corresponds approximately to a
confidence region of $\sim68\,\%$ of including the true value of the
two parameters of interest.

\section{Summary}
As we enter a phase of high-precision cosmological measurements,
sophisticated tools for data analysis are required, especially to
evaluate potential systematic effects related to the method. The
Monte-Carlo simulation package SNOC is mainly designed for the study
of the precision and possible biases of measured cosmological
parameters from high-$z$ Type Ia SNe. Besides calculating the
luminosity distance for arbitrary cosmological parameters: $H_0,
\Omega_M, \Omega_X, \alpha_X(z)$, the code also allows for the
estimates of observed magnitudes for high-$z$ SNe taking into account
the inhomogeneity of matter and dimming by, e.g., dust or more exotic
process such as photon-axion oscillations.  Along with the simulation,
a maximum-likelihood analysis package has been developed for fitting
of cosmological parameters and possible bias.


\section*{Acknowledgements}
The authors would like to thank Ch.~Walck for providing us
with his random number generator and  P.~Nugent for
the spectral template for Type Ia supernovae.
J.~Edsj\"o, M.~Eriksson, C.~Gunnarsson
and S.~Perlmutter are acknowledged for useful discussions. AG is
a Royal Swedish Academy Research Fellow supported by a grant from the
Knut and Alice Wallenberg Foundation.


\onecolumn
\section*{Appendix A: Summary of SNOC input parameters} 
\subsection*{The Input for the Monte-Carlo tool}
\begin{center}
\begin{longtable}{ll} 
    parameter name & description \\
    \hline\hline
{\tt NUMSNE}   & number of SNe to be generated                                             \\ 
{\tt SEED}     & seed for random number generator (if negative use computer clock)                \\ 
{\tt SNTYPE}   & SN type                                                                   \\ 
    \hline\hline
    \multicolumn{2}{c}{Redshift distribution} \\ \hline
{\tt GENDIS}   & functional form of SN redshift distribution option                              \\  
{\tt MEANZ}    & mean redshift of SN sample (for Gaussian distribution only)                      \\ 
{\tt SIGMAZ}   & $\sigma$ in redshift  of SN sample (for Gaussian distribution only)              \\ 
{\tt MAXZ}     & maximum SN redshift                                                       \\ 
{\tt MINZ}     & minimum SN redshift                                                       \\ 
{\tt ZFILE}    & name of file with tabulated $z$-distribution of SNe (optional)                     \\ 
{\tt TAUIA}    & $\tau$-parameter to in SFR redshift distribution \citep{DahlenFransson}             \\ 
    \hline\hline
    \multicolumn{2}{c}{Lightcurve} \\ \hline
{\tt INTSIG}   & intrinsic magnitude spread of SNe                                         \\ 
{\tt BUILDLC}  & option to build SN lightcurves                                                   \\   
{\tt LCPOINT}  & number of points in lightcurve (optional)                                        \\ 
{\tt LCNOISE}  & measurement error in each lightcurve point                                       \\ 
{\tt LCSPACE}  & number of days between lightcurve points                                         \\ 
{\tt LCTMIN}   & earliest day in lightcurve                                                       \\ 
{\tt FILTER}    & default filter for lightcurve (BVRIJ)                                            \\ 
{\tt USFILT}    & user specified filters                                                           \\ 
{\tt STRETCH}   & lightcurve ``stretch'', relevant for Type Ia SNe \citep{perlmutter97}             \\ 
{\tt STRCORR}   & lightcurve ``stretch-brightness'' correlation \citep{perlmutter97}                 \\ 
{\tt STRSIG}    & intrinsic lightcurve ``stretch'' standard deviation                              \\
     \hline\hline
    \multicolumn{2}{c}{Cosmological parameters} \\ \hline
{\tt HUBBLE}    & Hubble parameter, $h$ ($H_0=h\cdot$100 km s$^{-1}$ Mpc$^{-1}$)                   \\ 
{\tt OMEGAM}    & mass density, $\Omega_M(z=0)$                                                    \\ 
{\tt OMEGAX}    & dark energy, $\Omega_X(z=0)$                                                     \\ 
{\tt EOSTYPE}   & dark energy equation of state form (constant or time varying)                    \\   
{\tt ALPHAX0}   & dark energy equation of state parameter, $w_0 = p_X/\rho_X$ at $z$=0               \\ 
{\tt ALPHAX1}   & dark energy equation of state parameter, first derivative, $w_1 = {dw \over dz}$\\ 
{\tt EOSFILE}   & dark energy equation of state parameter, tabulated file with $w(z)$ (optional)   \\
    \hline\hline
    \multicolumn{2}{c}{Gravitational lensing} \\ \hline 
{\tt LENS}      & parametrization of matter lenses (compact objects, uniform, SIS or NFW)                   \\ 
{\tt DOFRAC}    & option to use alternating lens types                                              \\  
{\tt FRACHOM}   & fraction of path in homogeneous matter                                           \\ 
{\tt FRACPOI}   & fraction of path in an environment of compact objects                            \\ 
{\tt FRACUNI}   & fraction of path in an environment of uniform objects                            \\ 
{\tt FRACISO}   & fraction of path in an environment of SIS mass distributions                     \\ 
{\tt FRACNFW}   & fraction of path in an environment of NFW mass distributions                     \\  
{\tt CELLSZ}    & initial cell size [Mpc] (when MTYPE 1)                                           \\ 
{\tt MATTERB}   & size of homogeneous matter ball [Mpc] (when LENS 'uni')                          \\ 
{\tt MULTI}     & option to treat multiple images of gravitationally lensed SNe             \\ 
{\tt REJTYPE}   & event rejection level for secondary (lensed) images                              \\
{\tt CLUSTER}   & option to simulate lensing due to foreground galaxy cluster                       \\ 
{\tt ZCL}       & redshift of foreground galaxy cluster                                             \\ 
{\tt VELDISP}   & dispersion velocity of foreground galaxy cluster                                  \\ 
{\tt MASSCLU}   & mass of foreground galaxy cluster                                                 \\ 
{\tt RADCLU}    & radius of foreground cluster image field                                              \\ 
    \hline\hline
    \multicolumn{2}{c}{Dust extinction} \\ \hline 
{\tt DOGREY}     & option to add intergalactic dust in the line-of-sight                            \\ 
{\tt MINWAVE}    & shortest observable wavelength                                                   \\ 
{\tt MAXWAVE}   & longest  observable wavelength                                                    \\ 
{\tt NUMPOI}    & number of points in the wavelength range                                          \\ 
{\tt GREYTAB}   & option to use tabulated grey dust parameters vs redshift                          \\ 
{\tt GDFILE}    & file with grey dust tabulated data                                                \\ 
{\tt RV  }      & intergalactic dust parameter, see Eq.~(\ref{eq:rv}) \\

{\tt GREYSC}    & mean free path for intergalactic dust scattering \\ 

{\tt ZLIM}       & turning redshift in dust model, see Eq.~(\ref{eq:dustpara})                        \\  
{\tt P1}         & exponent in dust density redshift evolution below ZLIM, see Eq.~(\ref{eq:dustpara})    \\  
{\tt P2}        & exponent in dust density redshift evolution above ZLIM, see Eq.~(\ref{eq:dustpara})     \\
{\tt DOHOST}     & option to simulate host galaxy type, brightness and dust extinction              \\ 
{\tt RVHOST}    & $R_V$ for host galaxy dust                                                        \\ 
{\tt HOSTSC}    & mean free path for host galaxy dust scattering                                    \\ 
{\tt DOGLXY}     & option to consider dust extinction from foreground galaxies                      \\ 
{\tt RVGLXY}    & $R_V$ for foreground galaxy dust                                                  \\ 
{\tt GLXYSC}    & mean free path for foreground galaxy dust scattering                      \\   
    \hline\hline
    \multicolumn{2}{c}{Photon-axion oscillations} \\ \hline 
{\tt DOMIX}    & option to consider photon-axion oscillations                         \\ 
{\tt B0}       & mean comoving magnetic field  \\
{\tt B0WID}    & width of uniform dispersion of magnetic field \\
{\tt MMA}      & strength of photon-axion coupling \\
{\tt MA}       & axion mass \\
{\tt NE}       & mean comoving electron density \\
{\tt NEWID}    & width of uniform dispersion of electron density \\
    \hline\hline
    \multicolumn{2}{c}{K-corrections} \\ \hline
{\tt DOKCORR}   & calculate K-corrections at maximum from restframe B to observed BVRIJ            \\  
{\tt KFULL}     & write out magnitude distortions from host galaxy dust or line-of-sight dust          \\ 
    \hline\hline
    \multicolumn{2}{c}{Galaxy parameters} \\ \hline
{\tt EVOLUM}    & evolution if host galaxy luminosities according to  $L(z)=L(0)(1+z)^{\rm EVOLUM}$ \\  
{\tt MTYPE}     & equal mass galaxies or weighted by Schechter function                             \\ 
{\tt SCHEAL}    & Schechter function parameter                                              \\ 
    \hline\hline
    \multicolumn{2}{c}{SN search} \\ \hline
{\tt DOSEAR}    & option to simulate search conditions                                      \\ 
{\tt TIMEGAP}   & time between reference and discovery images in
    search mode                        \\
\caption{The SNOC Monte-Carlo tool steering parameters}
\label{tab:snoc_input}
\end{longtable}
\end{center}

\subsection*{The Input for the Analysis Tool}
\begin{center}
\begin{longtable}{l p{10cm}} 
    parameter name & description \\
    \hline\hline
    {\tt nrsne} & number of SNe to use in the fit\\
    {\tt zrange} & only use SNe within specified $z$-range\\
    {\tt nrparsigma} & the number of parameters that should
                       simultaneously lie within a specific deviation range.\\
    {\tt flat} & assume that the Universe is flat\\
    {\tt noflat} & do not assume that the Universe is flat\\
    \hline\hline
    \multicolumn{2}{c}{Fit parameters} \\ \hline
    \multicolumn{2}{p{13cm}}{These parameters specify which of the
      cosmological parameters that should be fixed and which that
      should be fitted. Each line consist of a parameter name and four
      numerical values, \texttt{s}, \texttt{e}, \texttt{g} and
      \texttt{c}. For a parameter to be fitted \texttt{s} and
      \texttt{e} cannot be identical, and \texttt{g} should be an
      integer greater than 1. For a grid search \texttt{s} and
      \texttt{e} specifies the grid interval for each parameter and
      \texttt{g} then number of steps. The value \texttt{c} should be
      the correct value of each parameter that was used in the
      Monte-Carlo. For the iterative minimization methods a start
      value will randomly be selected between \texttt{s} and \texttt{e}.}\\
    {\tt w\_0     s e g c} & non-redshift dependent part of
                             the equation of state parameter\\
    {\tt w\_1     s e g c} & linearly redshift dependent part of
                             the equation of state parameter\\
    {\tt omega\_M s e g c} & mass density, $\Omega_M$ ($z=0$)\\
    {\tt omega\_X s e g c} & dark energy, $\Omega_X$ ($z=0$)\\
    {\tt mscript  s e g c} & intercept in the Hubble diagram, $\sm$,
                             only used for Hubble diagram fits \\
    {\tt hubble0  s e g c} & Hubble parameter, $h$ ($H_0 =
                             h\cdot100$~km~s$^{-1}$ Mpc$^{-1}$), only used
                             for fits of multiply imaged objects\\
    \hline\hline
    \multicolumn{2}{c}{Nuisance parameters} \\ \hline
    {\tt NUISM} & fit $\sm$ as a nuisance parameter\\ 
    {\tt NUIFP} & fit the fraction of compact objects in the Universe\\   
    \hline\hline
    \multicolumn{2}{c}{Type Ia fits} \\ \hline
    {\tt gausserror} & the uncertainty in the magnitude $\sigma$\\
    {\tt addintsig} & add a dispersion to the measured magnitudes\\
    {\tt nointsig} & do not add a dispersion to the measured magnitudes\\
    {\tt addlenseffect} & add the magnitude dispersion due to lensing \\
    {\tt nolenseffect} & do not add the magnitude dispersion due to lensing \\
    {\tt magresiduals} & calculate the magnitude residuals
                         from the best cosmology fit\\
    {\tt nomagresiduals} & do not calculate the magnitude residuals
                           from the best cosmology fit\\
    {\tt gauss} & assume Gaussian magnitude distribution\\
    {\tt gausslin} & assume a modified Gaussian magnitude distribution
                     to handle gravitational lensing\\
    \hline\hline
    \multicolumn{2}{c}{Multiple image fits} \\ \hline
    {\tt domulti} & do a multiple image fit instead of the default
                  Type Ia fit\\ 
    {\tt ibandmax} & maximum allowed I-band magnitude for the SNe\\ 
    {\tt lenstype} & which lens model to use, SIS or POI\\
    {\tt etimedelay} & uncertainty in the time delay in days\\   
    {\tt eimsep} & uncertainty in the image separation in arcsecs\\
    {\tt imrat} & uncertainty in the flux ratio\\
    {\tt ezl} & uncertainty on lensing galaxy redshift\\
    {\tt emodel} & systematic error on lensing model\\
    \hline\hline
    \multicolumn{2}{c}{Minimization method} \\ \hline 
    {\tt davidonmin} & use the Davidon Variance Algorithm\\ 
    {\tt nodavidonmin} & do not use the Davidon Variance Algorithm\\ 
    {\tt gridsearch} & use the grid search\\
    {\tt nogridsearch} & do not use the grid search\\
    {\tt powellmin} & use Powell's Quadratically Convergent Method\\ 
    {\tt nopowellmin} & do not use Powell's Quadratically Convergent Method\\ 
    \caption{SNOC analysis steering parameters}
\label{tab:snalys_input}
\end{longtable}
\end{center}

\section*{Appendix B:The SNOC Output}
\subsection*{The Output of the Monte-Carlo Tool}
SNOC produces an ASCII output file that can be easily interpreted and
piped into histogramming tools.\footnote{We use the CERNLIB package
PAW for that purpose} Table \ref{tab:snoc_output} shows the first
event record of a SNOC simulation.

\begin{center}
{\large SNOC output file: example} 
\begin{longtable}{p{1.8cm}p{5.3cm}|l}
variable & value & description \\ \hline \hline
{\tt event}   &  1           & event number \\ 
{\tt sntype}   &    'Ia  '    & SN type \\ 
{\tt hubble}   &    0.6500    & Hubble parameter \\ 
{\tt omegam}   &   0.2800     & $\Omega_M(z=0)$ \\ 
{\tt omegax}   &   0.7200     &  $\Omega_X(z=0)$ \\ 
{\tt alphax0}  &   -1.0000    &  $w_0(z=0)$ \\ 
{\tt Ntotal}   &       237    &  number of SNe sq.deg$^{-1}$year$^{-1}$ \\
{\tt nzbins}   &      28      &  number of $z$-bins for rates \\ 
{\tt zcent}    &    0.125     0.175   0.225  0.275   0.325  0.375   0.425  0.475   0.525  0.575 
             0.625     0.675      0.725     0.775      0.825     0.875     0.925     0.975  
             1.025     1.075     1.125     1.175     1.225     1.275     1.325     1.375  
             1.425     1.475 & center of $z$-bin \\ 
{\tt zrate}    &    0.749  1.414  2.154  2.770  3.741  4.576   5.076  6.144  6.896  7.161 
             8.218     8.818     8.828     9.824    10.260    10.043    10.961    11.250 
            10.850    11.688    11.856    11.319   12.083    12.156   11.523    12.223 
            12.223    11.527 & differential SN rate  \\ 
{\tt zs}       &      0.4318 & redshift of SN (in event \#1) \\ 
{\tt lookb}    &      0.3264 & lookback time (Hubble time units)\\ 
{\tt lookGyr}  &      4.9234 &  lookback time in Giga-years\\ 
{\tt dL}       &      0.5586 & luminosity distance in units of ${c \over H_0}$\\ 
{\tt BZmag}    &     22.9903 &  effective B-band magnitude of event \\
{\tt intsig}   &      0.2662 &  intrinsic B-band offset from mean B-band Ia magnitude \\ 
{\tt emptyb}   &  1.0241 &  luminosity ratio empty beam/filled beam\\ 
{\tt emptydm}  &   0.0259 &  magnitude offset empty beam - filled beam \\
{\tt lensrat}  &   1.0212 &  lensing flux ratio demagnification \\ 
{\tt linsdm}   &  0.0228 &  lensing demagnification (magnitude)\\ 
{\tt ncells}   &        386 &  number of galaxy cells along path  \\ 
{\tt ncaustics}  &        0 &  number of parity-inversions of image along path \\ 
{\tt nreject}   &         0 &  number of image rejections before keeping event \\ 
{\tt ttot}      & 1508.5641 &  travel time (Mpc) \\ 
{\tt galtype}   &   'Sbc'  &  host galaxy type \\ 
{\tt galabsB}   & -16.4987 &  host galaxy absolute magnitude \\ 
{\tt galBband}  &  26.2917 &  host galaxy B-band magnitude \\ 
{\tt galVband}  &  25.0382 &  host galaxy V-band magnitude \\ 
{\tt galRband}  &  24.2002 &  host galaxy R-band magnitude \\ 
{\tt galIband}  &  23.5455 &  host galaxy I-band magnitude \\ 
{\tt hostdm}    & 0.0000  &   effective B-band extinction in host \\ 
{\tt greysc}    &58200.000 &  intergalactic dust mean free path (Mpc) \\ 
{\tt r\_v}       &   9.5000 &  $R_V$ intergalactic dust \\ 
{\tt zgreylim}  &   0.5000 &  intergalactic dust, density turn-point, see 
Eq.~(\ref{eq:dustpara})\\ 
{\tt greyexp1}  &   3.0000 &  intergalactic dust, exponent below turn-point, see Eq.~(\ref{eq:dustpara})\\ 
{\tt greyexp2}  &   0.0000 &  intergalactic dust, exponent above turn-point, see Eq.~(\ref{eq:dustpara})\\ 
{\tt greydm}    & 0.5118 &  intergalactic dust magnitude extinction of effective B-band \\ 
{\tt numgrey}   &       30 &  number of wavelength bins for intergalactic reddening  \\ 
{\tt lambdagr}  &   4000.0    4448.3    4896.6   5344.8    5793.1    6241.4
            6689.7    7137.9     7586.2    8034.5    8482.8    8931.0
            9379.3    9827.6   10275.9   10724.1   11172.4   11620.7 
           12069.0   12517.2   12965.5   13413.8   13862.1   14310.3 
           14758.6   15206.9  15655.2   16103.4   16551.7   17000.0  & wavelength central value \\
{\tt greyvec}  & 
0.539E-01 0.558E-01 0.560E-01 0.548E-01 0.532E-01 0.514E-01
         0.496E-01 0.476E-01 0.456E-01 0.436E-01 0.415E-01 0.395E-01
         0.375E-01 0.356E-01 0.337E-01 0.318E-01 0.300E-01 0.283E-01
         0.267E-01 0.252E-01 0.239E-01 0.226E-01 0.214E-01 0.204E-01
         0.194E-01 0.185E-01 0.176E-01 0.168E-01 0.161E-01 0.154E-01  & differential extinction \\
{\tt mixdm}       & 0.2443  & effective B-band extinction due to photon-axion oscillations\\ 
{\tt KcorBB0}     & 1.2692  &  K$_{\rm BB}$-correction, LC maximum  \\ 
{\tt KcorBV0}     & -0.1974 &  K$_{\rm BV}$-correction, LC maximum  \\ 
{\tt KcorBR0}     & -0.7003 &  K$_{\rm BR}$-correction, LC maximum  \\
{\tt KcorBI0}     & -0.7416 & K$_{\rm BI}$-correction, LC maximum  \\ 
{\tt KcorBJ0}     & -0.1928 &  K$_{\rm BJ}$-correction, LC maximum  \\ 
{\tt KcorBB1}     & 1.2808  & K$_{\rm BB}$-correction (includes host galaxy extinction)   \\ 
{\tt KcorBV1}     & -0.1873 & K$_{\rm BV}$-correction (includes host galaxy extinction)   \\
{\tt KcorBR1}     & -0.6916 & K$_{\rm BR}$-correction (includes host galaxy extinction)   \\
{\tt KcorBI1}     & -0.7349 & K$_{\rm BI}$-correction (includes host galaxy extinction)   \\
{\tt KcorBJ1}     & -0.1892 & K$_{\rm BJ}$-correction (includes host galaxy extinction)   \\
{\tt KcorBB2}     & 1.2692  & K$_{\rm BB}$-correction (includes foreground galaxy extinction)   \\
{\tt KcorBV2}     & -0.1974 & K$_{\rm BV}$-correction (includes foreground galaxy extinction)   \\
{\tt KcorBR2}     & -0.7003 & K$_{\rm BR}$-correction (includes foreground galaxy extinction)   \\
{\tt KcorBI2}     & -0.7416 & K$_{\rm BI}$-correction (includes foreground galaxy extinction)   \\
{\tt KcorBJ2}     & -0.1928 & K$_{\rm BJ}$-correction (includes foreground galaxy extinction)   \\
{\tt KcorBB3}     & 1.3246  & K$_{\rm BB}$-correction (includes intergalactic extinction)   \\
{\tt KcorBV3}     & -0.1434 & K$_{\rm BV}$-correction (includes intergalactic extinction)   \\
{\tt KcorBR3}     & -0.6500 & K$_{\rm BR}$-correction (includes intergalactic extinction)   \\
{\tt KcorBI3}     & -0.6978 & K$_{\rm BI}$-correction (includes intergalactic extinction)   \\
{\tt KcorBJ3}     & -0.1665 & K$_{\rm BJ}$-correction (includes intergalactic extinction)   \\
{\tt KcorBB4}     & 1.4405  & K$_{\rm BB}$-correction (includes photon-axion oscillation)   \\
{\tt KcorBV4}     & -0.0273 & K$_{\rm BV}$-correction (includes photon-axion oscillation)   \\
{\tt KcorBR4}     & -0.4390 & K$_{\rm BR}$-correction (includes photon-axion oscillation)   \\
{\tt KcorBI4}     & -0.4464 & K$_{\rm BI}$-correction (includes photon-axion oscillation)   \\
{\tt KcorBJ4}     & 0.1988  & K$_{\rm BJ}$-correction (includes photon-axion oscillation)   \\
{\tt KcorBB}      & 1.5257  & K$_{\rm BB}$-correction total extinction folded   \\ 
{\tt KcorBV}      & 0.0546  & K$_{\rm BV}$-correction total extinction folded   \\ 
{\tt KcorBR}      & -0.3646 & K$_{\rm BR}$-correction total extinction folded   \\ 
{\tt KcorBI}      & -0.3843 & K$_{\rm BI}$-correction total extinction folded   \\ 
{\tt KcorBJ}      & 0.2350  & K$_{\rm BJ}$-correction total extinction folded   \\  
{\tt stretch}     &0.9457 &  lightcurve stretch \\ 
{\tt strcorr}     &0.0543 &  stretch-brightness correction \\ 
{\tt npoints}     &     8 &  number of lightcurve points \\ 
{\tt lcdate}      &-13.54  -6.77   0.00  6.77   13.54  20.31  27.08 33.85 & dates from maximum \\
{\tt lcmag}       &    0.42  -0.05  -0.58  -0.28   0.13   0.54   1.15   1.61 
                  & m$_R(t)$ - {\tt BZmag} (R-band lightcurve) \\
{\tt lcnoise}     & 0.1000  & applied noise standard deviation in LC points \\ 
{\tt dayref}     & -12.1837 & days from maximum for reference image (rest frame) \\ 
{\tt daynew}     &  3.1770 & days from maximum for search image (rest frame) \\ 
{\tt magdiffB}   & -0.9194 & SN B-band magnitude difference new - ref epoch \\ 
{\tt magdiffV}   & -1.2900 & SN V-band magnitude difference new - ref epoch \\ 
{\tt magdiffR}   & -1.4338 & SN R-band magnitude difference new - ref epoch \\ 
{\tt magdiffI}   & -1.5398 & SN I-band magnitude difference new - ref epoch \\ 
{\tt magdiffJ}   & -1.3374 & SN J-band magnitude difference new - ref epoch \\ 
{\tt magnewB}    &  1.5924 & SN B-band magnitude (-BZmag) at discovery epoch \\ 
{\tt magnewV}    & -0.0228 & SN V-band magnitude (-BZmag) at discovery epoch \\ 
{\tt magnewR}   & -0.6042 & SN R-band magnitude (-BZmag) at discovery epoch \\ 
{\tt magnewI}    & -0.7418 & SN I-band magnitude (-BZmag) at discovery epoch \\ 
{\tt magnewJ}   & -0.0819 & SN J-band magnitude (-BZmag) at discovery epoch \\ 
{\tt end} &  & end of event \# 1 \\ 
\caption{Extract of SNOC output showing the first event record}
\label{tab:snoc_output}
\end{longtable}
\end{center}

\subsection*{The Output of the Analysis Tool}\label{sec:snalysout}
\begin{center}
  {\large SNOC output file: example}
\end{center}
\begin{verbatim}
# THIS IS THE OUTPUT GENERATED BY SNALYS
# ======================================
# The number of supernovae used:    2300
# The number of parameters estimated:  2

# Flat universe assumption: no
# Execution started at: Wed Mar  6 16:24:15 2002
# Total execution time: 00:01:33.00
# Used CPU-time: total = 00:01:35.43
#                user = 00:01:35.34
#                system = 00:00:00.07

# Powell Minimization
# ===================
# Start values
omega_m     0.333
omega_x     0.733
# The min of the ML function (powell):      0.010
# Cosmology (powell)
omega_m     0.299860
omega_x     0.699683

# Davidon Minimization
# ====================
# Start values 
omega_m     0.300
omega_x     0.700
# The min of the ML function (davidon):      0.010
# Cosmology (davidon)
omega_m    0.299983 +/-  0.023836
omega_x    0.699974 +/-  0.046131
# Correlation matrix (davidon)   
           omega_m    omega_x    
omega_m      1.00000   0.95052   
omega_x      0.95052   1.00000   

# Grid Search
# ===========
# The min of the ML function (grid):      6.729
# Cosmology (grid)
omega_m    0.333333 +   0.33E-01 -   0.67E-01
omega_x    0.733333 +   0.00E+00 -   0.67E-01
# Conditions for grid search.
# Parameter      start   stopnr.step
     omega_m    0.10 0.40   10
     omega_x    0.40 1.00   10
# The ML-function from Grid Search
#omega_m   omega_x   
  82.4738235
  42.1675644
  51.1707954
  ...
\end{verbatim}

\end{document}